\def\year{2022}\relax

\documentclass[letterpaper]{article} 
\usepackage{aaai22}  
\usepackage{times}  
\usepackage{helvet}  
\usepackage{courier}  
\usepackage[hyphens]{url}  
\usepackage{graphicx} 
\usepackage{natbib}  
\usepackage{caption} 
\DeclareCaptionStyle{ruled}{labelfont=normalfont,labelsep=colon,strut=off} 
\usepackage{newfloat}
\usepackage{listings}
\usepackage{cite}
\usepackage{amsmath,amssymb,amsfonts}
\usepackage[dvipsnames,table,xcdraw]{xcolor} 
\usepackage{booktabs}
\usepackage{multirow}
\usepackage{textcomp}

\def\BibTeX{{\rm B\kern-.05em{\sc i\kern-.025em b}\kern-.08em
    T\kern-.1667em\lower.7ex\hbox{E}\kern-.125emX}}
\begin{document}

\title{Fine-Grained Classroom Activity Detection from Audio with Neural Networks}
\author{
    Eric Slyman\textsuperscript{\rm 1,2},
    Chris Daw\textsuperscript{\rm 3},
    Morgan Skrabut\textsuperscript{\rm 3}
    Ana Usenko\textsuperscript{\rm 2},
    Brian Hutchinson\textsuperscript{\rm 2,4}%
    \thanks{The authors gratefully acknowledge the NVIDIA Corporation for the donation of GPUs used in this work. All of the authors were at WWU when the work was conducted.}
}
\affiliations{
    
    \textsuperscript{\rm 1}Oregon State University, Corvallis, OR 97330, USA\\
    \textsuperscript{\rm 2}Pacific Northwest National Laboratory, Seattle, WA 98109 USA\\
    \textsuperscript{\rm 3}DocuSign, Seattle, WA 98104, USA\\
    \textsuperscript{\rm 4}Western Washington University, Bellingham, WA 98225, USA\\
    slymane@oregonstate.edu,
    \{chris.daw,morgan.skrabut\}@docusign.com
    \{eric.slyman,brian.hutchinson,ana.usenko\}@pnnl.gov,
    brian.hutchinson@wwu.edu,
}

\maketitle

\begin{abstract}
Instructors are increasingly incorporating student-centered learning techniques in their classrooms to improve learning outcomes. In addition to lecture, these class sessions involve forms of individual and group work, and greater rates of student-instructor interaction. Quantifying classroom activity is a key element of accelerating the evaluation and refinement of innovative teaching practices, but manual annotation does not scale. In this manuscript, we present advances to the young application area of automatic classroom activity detection from audio. Using a university classroom corpus with nine activity labels (e.g., ``lecture,'' ``group work,'' ``student question''), we propose and evaluate deep fully connected, convolutional, and recurrent neural network architectures, comparing the performance of mel-filterbank, OpenSmile, and self-supervised acoustic features. We compare 9-way classification performance with 5-way and 4-way simplifications of the task and assess two types of generalization: (1) new class sessions from previously seen instructors, and (2) previously unseen instructors. We obtain strong results on the new fine-grained task and state-of-the-art on the 4-way task: our best model obtains frame-level error rates of 6.2\%, 7.7\% and 28.0\% when generalizing to unseen instructors for the 4-way, 5-way, and 9-way classification tasks, respectively (relative reductions of 35.4\%, 48.3\% and 21.6\% over a strong baseline). When estimating the aggregate time spent on classroom activities, our average root mean squared error is 1.64 minutes per class session, a 54.9\% relative reduction over the baseline.
\end{abstract}

\section{Introduction}
Meeting society's needs for graduates skilled in science, technology, engineering, and mathematics (STEM) is of the most important challenges universities face. Innovative student-centered and active learning pedagogical strategies are key tools in addressing these challenges, since they increase retention rates of STEM students and enhance learning outcomes~\cite{olson2012engage,freeman2014active}. Adopting these strategies requires diversifying the use of class time (e.g., opportunities for small group work and greater rates of student-instructor interaction). Instructors need to be able to quantify, analyze, and reflect on their use of class time on student-centered activities to most effectively maximize student learning outcomes. Traditionally, knowing how class time is spent requires bringing in a trained observer to hand-annotate during lecture, for example, using one of the popular classroom annotation schemes~\cite{smith2013classroom,Sawada2002MeasuringRP,Velasco2016CharacterizingIP,Eddy2015PORTAALAC}. Manual annotation, however, cannot scale to meet the need for accurate classroom activity annotation and therefore an automated approach is needed.

In early work, Wang et al.~\shortcite{wang2014automatic} propose training a random forest over the outputs of a modified proprietary wearable device, ``LENA,''~\cite{ford2008lena} to predict ``teacher lecturing,'' ``whole class discussion,'' and ``student group work.'' Owens et al. address the problem of detecting activities in college-level lecture audio using portable audio recording devices~\shortcite{Owens3085}. Their DART system uses binary decision trees with amplitude statistics as features to classify a 4-way label set: ``single voice,'' ``multi voice,'' ``no voice'' and ``other''. Cosbey et al. improve upon DART using deep and recurrent neural network approaches and mel-filterbank features ~\shortcite{cosbey2019deep}. They demonstrate substantial reductions in error on the same data with a gated recurrent unit (GRU) neural network, reporting 31.7\% and 45.1\% reduction in frame-level error over DART for new class sessions from previously seen instructors and new class sessions from previously unseen instructors, respectively. 

Li et al. focus on the task of distinguishing teacher from student speaker roles using siamese networks and instructor voice enrollment~\shortcite{Li2020SiameseNN}. In separate work, Li et al. use attention over automatic speech recognition transcripts to improve performance over models using acoustic features on the same task~\shortcite{li2020multimodal}. Donnelly et al. report experiments using close talking noise-canceled microphone data collected from middle school classrooms~\shortcite{donnelly2016automatic}. Of the 17 activity labels, they aim to detect the most common five: ``question \& answer,'' ``procedures and directions,'' ``supervised seatwork,'' ``small group work,'' and ``lecture.''

Other work focuses on identifying specialized classroom activities such as detecting instructor questions~\cite{10.1145/3027385.3027417,10.1007/978-3-030-52237-7_22}, classifying student arguments into ``claims,'' ``evidence,'' and ``warrants''~\cite{lugini-litman-2018-argument}, or analyzing dialogic one-on-one instruction to detect activities such as ``greeting,'' ``guidance,'' and ``commending''~\cite{10.1007/978-3-030-52240-7_62}.

\begin{figure*}
  \centering
  \includegraphics[width=1\textwidth]{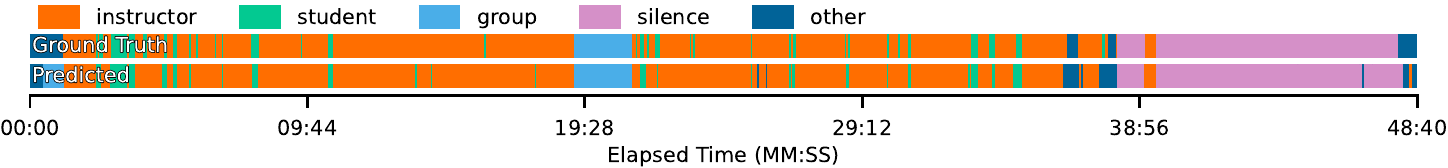}
  \caption{Activity trace of an unseen 49 minute class session from an instructor previously seen in training, representative of performance of our best model.}
  \label{fig:trace}
\end{figure*}

The work reported in this manuscript represents significant progress in classroom activity detection, as the first work to operate under the challenging, realistic scenario defined by the following: 
\begin{enumerate}
    \item {\bf Fine-grained classification.} We produce fine-grained (9-way) frame level predictions, e.g., distinguishing between instructor and student speech, and question from answer. To aid comparison with the existing literature, we also report 5-way and 4-way variants.
    \item {\bf Practical audio capture.} We only use audio recorded by commodity webcams at the instructor station. 
    \item {\bf Transcript free.} We do not rely on automatic speech recognition given the noisy and far-field environment.  
    \item {\bf Minimal or no ``enrollment.''} We do not assume any explicit enrollment data from students and assess generalization to previously unseen instructors and classrooms.
\end{enumerate}

To address these challenges we develop and thoroughly evaluate four deep learning models: deep fully connected, dilated temporal convolutional, and both unidirectional and bidirectional gated recurrent unit neural networks. 
We also compare the utility of three input feature sets: mel-filterbank features, OpenSmile features, and PASE+ self-supervised embeddings~\cite{ravanelli2020multi}.
We separately report generalization performance to new class sessions from both seen and unseen instructors. In addition to reporting the first fine-grained classroom activity results, we also obtain new state-of-the-art results on 4-way activity classification. Example model output is shown in Fig.~\ref{fig:trace}, using the 5-way classification scheme.
We have made the code for this work publicly available.\footnote{https://github.com/hutchresearch/fine-grained-cad}

\section{Methods} \label{sec:methods}

\subsection{Feature Extraction}
We consider three feature sets. When features from different sets are combined, we align frame centers and concatenate to produce a single feature vector per timestep.

\begin{enumerate}
\item {\it Mel-filterbank:}
We downsample the raw waveform to 16kHz and extract 40 log-scaled mel-filterbank features using the Python audio processing and analysis package librosa~\cite{librosa} with 500ms windows at 10ms offsets, standardizing each frequency bin independently using the training set statistics. Log mel-filterbank features have been widely used in recent years, in applications ranging from automatic speech recognition \cite{pratap2019wav2letter} to speaker verification \cite{wan2018generalized} and audio event detection \cite{fonseca2019learning}. Based on the performance reported in \cite{cosbey2019deep}, we use a significantly longer frame length than is common for these other applications.

\item {\it OpenSMILE:}
Using the open-source speech feature extraction toolkit, OpenSmile~\cite{opensmile2013}, we extract three frame-level features over 50ms frames at a 10ms offset: fundamental frequency, loudness, and voicing probability. We standardize each feature using the training set statistics. The first two features carry key prosodic information; for example, there are often distinctive differences in utterance-final pitch between declarative utterances and various question types. Patterns in the voicing probability may help distinguish between types of activities featuring multiple, single, or no speakers.

\item {\it PASE+ Embeddings:}
\cite{ravanelli2020multi} propose a problem-agnostic speech encoder (PASE+) for raw waveforms that is pretrained with multiple self-supervised tasks, such as predicting Mel-frequency cepstral coefficients and gammatone features, while adding artificial speech distortions to encourage robust representations. PASE features~\cite{Pascual2019} were shown to provide excellent performance for automatic speech recognition (ASR), speaker identification, and speaker emotion classification, before expanding to noise robust ASR in the PASE+ variant. We use the output of a pretrained PASE+, fine-tuned during training, as our third feature set to capture speaker information that is invariant to the noisy classroom environment. Waveforms are normalized before embedding 150ms frames at 10ms offsets.
\end{enumerate}

\subsection{Classifiers}
\begin{figure}
  \centering
  \includegraphics[width=\columnwidth]{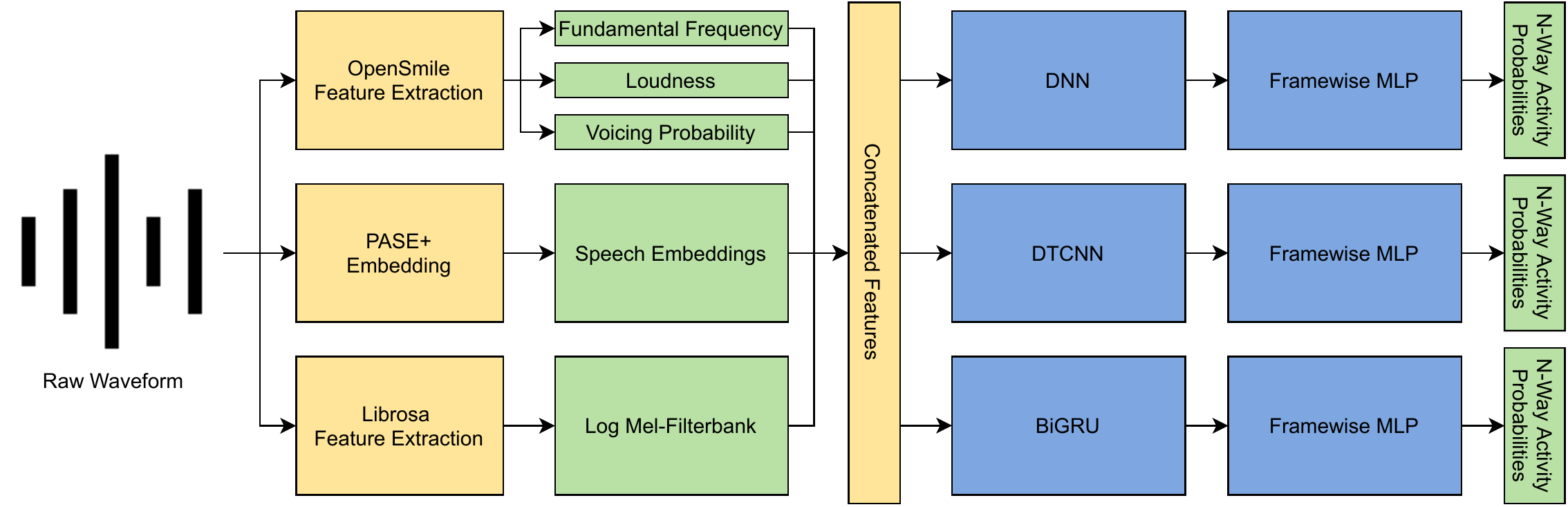}
  \caption{Summary of approach, from input waveforms to output n-way posterior probabilities over activity classes.}
  \label{fig:overview}
\end{figure}

We design and evaluate four classifiers based on the following deep learning architectures. In each case, we produce frame level posterior probabilities over the activity labels via an output framewise two-layer MLP with Leaky ReLU~\cite{maas2013rectifier} hidden activation and softmax output activation. The overarching network flow is shown in Fig.~\ref{fig:overview}.

\begin{enumerate}
\item {\it DNN:}
A fully connected, feed-forward Deep Neural Net (DNN). The input to the model is a context window of frame-level features. Specifically, let $x_t$ denote the input features of frame $t$, and $\hat{x}_t^K$ denote the feature-dimension concatenation of frames $x_{t-K},x_{t-K+1},\dots,x_{t-1},x_t,x_{t+1},\dots,x_{t+K}$, then our predictions are given by:
\begin{eqnarray}
    h_{t,0} & = & \text{LReLU}(W_0 \hat{x}_t^k + b_0) \\
    h_{t,1} & = & \text{LReLU}(W_1 h_{t,0} + b_1)     \\
            &\vdots &  \nonumber                      \\
    h_{t,n} & = & \text{LReLU}(W_n h_{t,n-1} + b_n)   \\
    y_t     & = & \text{MLP}(h_{t,n})
\end{eqnarray}
where each $W$ matrix and $b$ vector are learned model parameters, and LReLU denotes an element-wise Leaky ReLU non-linearity.

\begin{figure}
  \centering
  \includegraphics[width=0.4\textwidth]{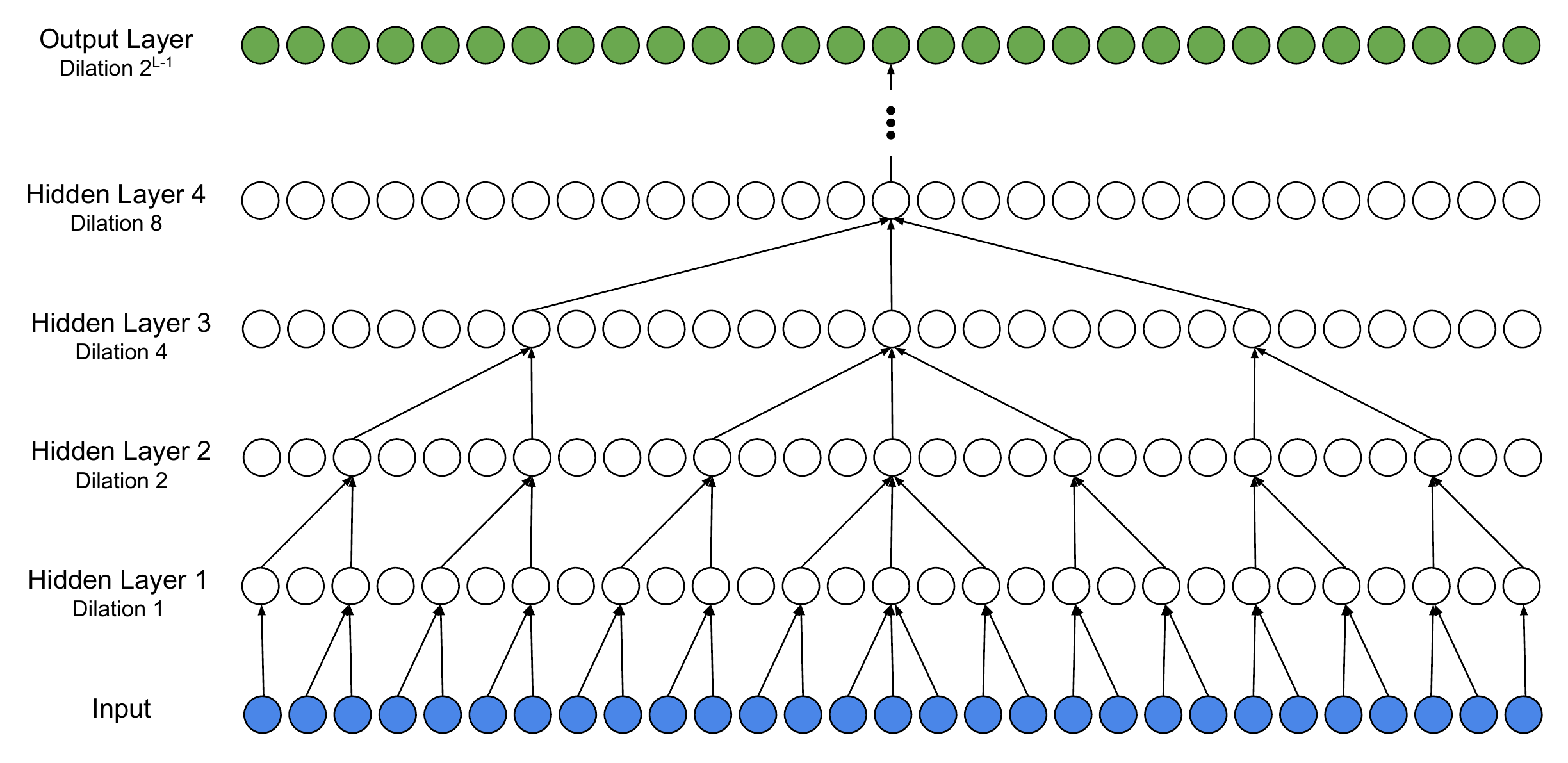}
  \caption{An example dilated temporal convolution model with dilation increasing by a factor of two each layer.}
  \label{fig:dilated}
\end{figure}

\item {\it DTCNN:}
A dilated temporal (1D) convolutional neural network (DTCNN), a non-causal variant of causal CNNs like WaveNet~\cite{oord2016wavenet}. The temporal dilation of the DTCNN provides the classifier with a significantly larger context around a given frame than standard convolutions. We expect this context benefits the classifier, especially near the boundary between two activities since the large receptive fields are symmetric around each prediction frame. The input to this model is first normalized with a 1-dimensional batch norm layer followed by dropout. Then the data is fed through $L$ layers of 1-dimensional dilated convolution. Our dilated convolution operation is illustrated in Fig.~\ref{fig:dilated}. The dilation factor for each layer is given by $b^{\ell-1}$, where $\ell$ is the CNN layer of the network and $b$ is a tuned hyperparameter. Each CNN layer is composed of a 1-dimensional convolution using the aforementioned dilation factor, leaky ReLU, and 1-dimensional batch norm.

\item {\it GRU:}
Given the significance of sequential information in audio, recurrent neural network architectures are an obvious fit. We train a gated recurrent unit (GRU) network~\cite{cho2014learning}, in order to implicitly condition predictions at time $t$ on all input features up to and including time $t$. Specifically, our activity prediction at time $t$, $y_t$, given the input $x_t$ is defined to be:
\begin{eqnarray}
    r_t & = & \sigma(W_{i,r}x_t + W_{h,r}h_{t-1} + b_{r}) \\
    z_t & = & \sigma(W_{i,z}x_t + W_{h,z}h_{t-1} + b_{z}) \\
    n_t & = & \text{tanh}(W_{i,n}x_t + b_{i,n} + \nonumber
    \\ && \hspace{3em} r_t \circ\ (W_{h,n}h_{t-1} + b_{h,n})) \\
    h_t & = & (1 - z_t) \circ n_t + z_t \circ h_{t-1} \\
    y_t & = & \text{MLP}(h_t)
\end{eqnarray}
where $r$, $z$, and $n$ are the reset, update, and new gates, $\sigma$ is the logistic sigmoid function, and $\circ$ is the Hadamard (element-wise) product. The matrices $W$ and vectors $b$ are learned model parameters. In the multi-layer case, the input of each layer is the post-dropout hidden state of the previous layer. The previous hidden state at the initial time step is a zero-vector.
Note that we also conducted experiments with Long Short-Term Memory (LSTM) networks \cite{hochreiter97}, but found no significant differences in performance as compared to the GRU, and thus do not report LSTM performance.
\item {\it BiGRU:}
Because our predictions are not made online, our system need not be causal: to predict the activity label at frame $t$ we are free to utilize acoustic cues in frames $t'>t$. Due to distinctive patterns of activity and interaction in a classroom future context may provide valuable cues to the current activity; as one example, observing a probable ``instructor answer'' in the near future increases the odds that the current activity is ``student question.'' We therefore experiment with a bidirectional GRU (BiGRU), which conditions on both past and future inputs. In a BiGRU, our prediction state at time $t$, $y_t$, given the input $x_t$ is defined to be:
\begin{eqnarray}
    \overrightarrow{h}_t & = & GRU(x_t, \overrightarrow{h}_{t-1}) \\
    \overleftarrow{h}_t  & = & GRU(x_t, \overleftarrow{h}_{t+1}) \\
    h_t                  & = & W_f \overrightarrow{h}_t + W_b \overleftarrow{h}_t + b \\
    y_t & = & \text{MLP}(h_t)
\end{eqnarray}
where all $\overrightarrow{h}_t$ denote the hidden states of the forward GRU (capturing information in the input from frames before $t$) and all $\overleftarrow{h}_t$ denote the hidden states of the backward GRU (capturing information in the input for frames after). In addition to the parameters of the forward and backward GRUs themselves, matrices $W_f$, $W_b$ and vector $b$ are also learnable parameters of the model.
\end{enumerate}

\subsection{Training and Tuning} \label{subsec:traintune}
We implement all architectures with the popular Python machine learning library PyTorch~\cite{paszke2019pytorch}. To track experimental results and hyperparameter tuning, we leverage Weights \& Biases~\cite{wandb}, a suite of centralized, web-based machine learning developer tools. To reduce the contribution to loss from easy sequences and increase it from underrepresented classes, we use class-balanced~\cite{cui2019class} focal loss~\cite{lin2017focal} with tuned $\beta$ and $\gamma$, reweighting the loss according to an estimated number of per-class samples then further reducing the contribution of well-classified samples to the final aggregated value. This formulation is found to be essential for preventing the network from overpredicting the ``single'' (sgl, 4-way), ``instructor'' (ist, 5-way), and ``instructor lecture'' (l, 9-way) majority class labels. Specifically, the class-balanced focal loss is:
\begin{eqnarray}
    \text{CB}_{\text{focal}}(p, y) = -\frac{1-\beta}{1-\beta^{n_y}}(1-p_y)^\gamma\ \text{log}(p_y).
\end{eqnarray}
Where $p_y$ is the model's posterior probability for target activity class $y$. Note that this differs from the sum of per-class logistic sigmoid activations used in the original work of~\cite{cui2019class} due to the mutually-exclusive nature of our classification task.

We optimize with AdamW~\cite{loshchilov2018decoupled}, a common adaptive gradient method with decoupled weight decay, and tune the learning rate and weight decay. Each network trains for up to 35 epochs using early stopping with a patience of nine epochs. We train over sequences of 3,000 frames for RNN-based classifiers and 12,000 frames for all others -- thirty seconds and two minutes of audio, respectively. Each minibatch is sampled randomly from all sequences in all class sessions, with minibatch size (number of sequences) tuned in $4$-$32$. Larger minibatch sizes require gradient accumulation to fit in GPU memory. To improve stability during training, we clip the L2-norm of the gradients to one and use a learning rate schedule in which the learning rate is halved every three epochs that development set performance does not improve. Training takes 10-20 hours on an Nvidia RTX 2080Ti GPU depending on configuration.

All above hyperparameters are tuned via Bayesian optimization~\cite{snoek}, a principled technique to select optimal hyperparameters for a black-box optimization function, independently for each classifier and labeling scheme combination in order to minimize development set error. Hyperband early stopping~\cite{li2017hyperband} is used to increase throughput in the number of hyperparameter choices we consider by selecting adaptive configurations as framed by a multi-armed bandit resource allocation problem, with four evaluation bands ranging from five to twenty epochs up to a maximum of thirty. For all classifiers, we tune the number of hidden layers, hidden size, and dropout rate. We tune the choice in normalization layers for DNN and DTCNN classifiers between affine and non-affine instance-norm~\cite{ulyanov2016instance}, batch-norm~\cite{ioffe2015batch}, and layer-norm~\cite{ba2016layern} which all serve to stabilize the loss function and improve training time.  Layer-norm performs best in all cases for the DTCNN, while instance-norm and layer-norm perform equally for the DNN. We tune the DTCNN dilation factor between $2$-$2.5$, with the best performance found at $2.1$, and apply dropout~\cite{srivastava2014dropout} immediately before passing the fused embeddings to the classifier and before the final two-layer MLP. When PASE+ embeddings are used, dropout is also applied to the raw audio waveform before generating an embedding.

\begin{table}
\caption{4-, 5- and 9-way Classroom activity labels.}
\label{tab:labels}
\centering
\resizebox{0.75\columnwidth}{!}{%
\begin{tabular}{r|c|c|c}
\toprule
\multicolumn{1}{c|}{\textbf{Activity}} & \textbf{4-Way} & \textbf{5-Way} & \textbf{9-Way} \\ \midrule
instructor announcement      & \multirow{6}{*}{sgl}     & \multirow{4}{*}{ist} & a     \\
instructor lecture           &                          &                      & l     \\
instructor asks question     &                          &                      & iq    \\
instructor answers question  &                          &                      & ia    \\ \cline{3-4}
students asks question       &                          & \multirow{2}{*}{stu} & sq    \\
students answers question    &                          &                      & sa    \\ \cline{2-4} 
group work                   & \multicolumn{2}{c|}{grp}                         & g    \\ \cline{2-4} 
silence                      & \multicolumn{2}{c|}{sil}                         & s    \\ \cline{2-4} 
other                        & \multicolumn{2}{c|}{oth}                         & o    \\ \bottomrule
\end{tabular}}
\end{table}

\subsection{Classroom Activity Labels}
Table~\ref{tab:labels} lists the nine labels used in our fine-grained label set. 
It also shows how our fine-grained labels are mapped to 5-way and 4-way simplifications of the task: 5-way distinguishes between student (stu) and instructor (ist), but not question and answer (iq, ia, sq, sa) or announcements (a) and lecture (l), while 4-way merges student and instructor into one ``single voice'' (sgl) category, as was used in~\cite{Owens3085,cosbey2019deep}. Note that ``other'' (oth) refers to unstructured, non-instructional time (e.g. before the class session begins). We train separate models for each activity label set.

\section{Experiments} \label{sec:experiments}
\subsection{Data}
We use a new dataset of classroom audio annotated with the 9-way label scheme (see Table~\ref{tab:labels}). The dataset consists of 58.7 hours of carefully hand-annotated classroom audio. All recorded class sessions are STEM classes at a master's granting university in the US Pacific Northwest, using commodity webcams placed at the instructor workstation, facing the students. It contains recorded class sessions from thirteen courses taught over two quarters by nine professors in eleven classrooms with widely varying size and layout. Occurrences and durations of all the labels in the dataset are enumerated in Table~\ref{tab:datatime}.

\begin{table}[]
\caption{Distribution of event types in data.}
\label{tab:datatime}
\centering
\begin{small}
\resizebox{0.75\columnwidth}{!}{%
\begin{tabular}{c|r|r|r}
\toprule
\multicolumn{1}{l|}{\textbf{Label}} & \multicolumn{1}{l|}{\textbf{Occurences}} & \multicolumn{1}{l|}{\textbf{Duration (min)}} & \multicolumn{1}{l}{\textbf{Percentage}} \\ \midrule
l     & 673 & 1427.59 & 40.0 \\ 
ia    & 1581& 656.72  & 18.4 \\ 
g     & 184 & 584.09  & 16.4 \\ 
a     & 325 & 230.09  &  6.4 \\ 
o     & 126 & 188.16  &  5.3 \\
iq    & 1202& 187.22  &  5.2 \\ 
s     & 118 & 115.58  &  3.2 \\ 
sq    & 766 & 92.47   &  2.6 \\ 
sa    & 960 & 89.32   &  2.5 \\ \hline
total & 5935& 3571.24 & 100.0 \\ \bottomrule
\end{tabular}}
\end{small}
\end{table}

Three of the professors are withheld to form Test 2 (6.4 hours), to test generalization of the trained system to entirely unseen instructors and classrooms.  Class sessions from the remaining six instructors are split into Train (35.8 hours), Development (8.5 hours), and Test 1 (8.1 hours). Due to IRB constraints, we are unable to publicly release this data.

\begin{table*}
    \caption{Experimental results, reporting mean average precision (mAP), weighted $F_1$ (F) and accuracy (Acc).  Lightness of cells corresponds to better performance by metric, with best scores in \textbf{bold}. Our two baselines are underlined; the second baseline (GRU + mel-filterbank) is the best configuration reported in \cite{cosbey2019deep}.}
\label{tab:main-results}
\centering
\small
\resizebox{\textwidth}{!}{%
\begin{tabular}{ll|ccc|ccc|ccc|ccc|ccc|ccc}
\toprule
 &  & \multicolumn{6}{c|}{\textbf{4-Way}}  & \multicolumn{6}{c|}{\textbf{5-Way}}  & \multicolumn{6}{c}{\textbf{9-Way}} \\
 &  & \multicolumn{3}{c|}{\textbf{Test 1}} & \multicolumn{3}{c|}{\textbf{Test 2}} 
    & \multicolumn{3}{c|}{\textbf{Test 1}} & \multicolumn{3}{c|}{\textbf{Test 2}} 
    & \multicolumn{3}{c|}{\textbf{Test 1}} & \multicolumn{3}{c}{\textbf{Test 2}} \\
\textbf{Model} & \textbf{Feature(s)} 
               & \textbf{mAP} & \textbf{F} & \textbf{Acc} 
               & \textbf{mAP} & \textbf{F} & \textbf{Acc} 
               & \textbf{mAP} & \textbf{F} & \textbf{Acc} 
               & \textbf{mAP} & \textbf{F} & \textbf{Acc} 
               & \textbf{mAP} & \textbf{F} & \textbf{Acc} 
               & \textbf{mAP} & \textbf{F} & \textbf{Acc} \\ \midrule
\multirow{1}{*}{}
 & \underline{Majority Class}      & .250 & .662 & .764 & .250 & .608 & .724 & .200 & .597 & .716 & .200 & .559 & .687 & .111 & .261 & .433 & .111 & .296 & .466 \\
\midrule
\multirow{7}{*}{DNN}
 & Mel-Filterbank      & \cellcolor[rgb]{1.00,0.64,0.44}.664 & \cellcolor[rgb]{0.99,0.86,0.62}.885 & \cellcolor[rgb]{0.99,0.88,0.63}.897 & \cellcolor[rgb]{0.99,0.57,0.40}.616 & \cellcolor[rgb]{1.00,0.71,0.49}.807 & \cellcolor[rgb]{1.00,0.73,0.51}.832 & \cellcolor[rgb]{0.96,0.42,0.36}.532 & \cellcolor[rgb]{1.00,0.73,0.51}.819 & \cellcolor[rgb]{1.00,0.78,0.54}.851 & \cellcolor[rgb]{0.93,0.35,0.37}.489 & \cellcolor[rgb]{1.00,0.63,0.43}.773 & \cellcolor[rgb]{1.00,0.68,0.46}.806 & \cellcolor[rgb]{0.62,0.18,0.50}.248 & \cellcolor[rgb]{0.55,0.16,0.51}.406 & \cellcolor[rgb]{0.70,0.21,0.48}.524 & \cellcolor[rgb]{0.55,0.16,0.51}.202 & \cellcolor[rgb]{0.59,0.17,0.50}.429 & \cellcolor[rgb]{0.72,0.21,0.48}.538 \\
 & OpenSmile             & \cellcolor[rgb]{1.00,0.64,0.44}.665 & \cellcolor[rgb]{1.00,0.81,0.57}.857 & \cellcolor[rgb]{1.00,0.83,0.59}.876 & \cellcolor[rgb]{0.99,0.55,0.39}.610 & \cellcolor[rgb]{1.00,0.83,0.59}.867 & \cellcolor[rgb]{0.99,0.86,0.61}.886 & \cellcolor[rgb]{0.98,0.47,0.36}.563 & \cellcolor[rgb]{1.00,0.63,0.43}.769 & \cellcolor[rgb]{1.00,0.71,0.49}.820 & \cellcolor[rgb]{0.92,0.34,0.38}.476 & \cellcolor[rgb]{1.00,0.68,0.47}.797 & \cellcolor[rgb]{1.00,0.74,0.51}.838 & \cellcolor[rgb]{0.72,0.21,0.48}.321 & \cellcolor[rgb]{0.60,0.18,0.50}.431 & \cellcolor[rgb]{0.73,0.22,0.47}.542 & \cellcolor[rgb]{0.64,0.19,0.49}.263 & \cellcolor[rgb]{0.77,0.23,0.46}.526 & \cellcolor[rgb]{0.87,0.29,0.41}.614 \\
 & PASE+               & \cellcolor[rgb]{0.96,0.40,0.36}.522 & \cellcolor[rgb]{1.00,0.81,0.57}.860 & \cellcolor[rgb]{1.00,0.83,0.59}.876 & \cellcolor[rgb]{0.94,0.37,0.37}.502 & \cellcolor[rgb]{1.00,0.68,0.47}.798 & \cellcolor[rgb]{1.00,0.64,0.44}.793 & \cellcolor[rgb]{0.90,0.32,0.39}.460 & \cellcolor[rgb]{1.00,0.63,0.43}.769 & \cellcolor[rgb]{1.00,0.68,0.47}.810 & \cellcolor[rgb]{0.87,0.29,0.41}.430 & \cellcolor[rgb]{1.00,0.64,0.44}.774 & \cellcolor[rgb]{1.00,0.65,0.45}.797 & \cellcolor[rgb]{0.66,0.20,0.49}.274 & \cellcolor[rgb]{0.58,0.17,0.50}.421 & \cellcolor[rgb]{0.70,0.21,0.48}.529 & \cellcolor[rgb]{0.56,0.16,0.51}.211 & \cellcolor[rgb]{0.65,0.19,0.49}.459 & \cellcolor[rgb]{0.70,0.21,0.48}.526 \\
 & Mel/OpenSmile         & \cellcolor[rgb]{0.99,0.88,0.64}.814 & \cellcolor[rgb]{0.99,0.94,0.70}.923 & \cellcolor[rgb]{0.99,0.93,0.68}.925 & \cellcolor[rgb]{1.00,0.74,0.51}.723 & \cellcolor[rgb]{0.99,0.88,0.63}.892 & \cellcolor[rgb]{0.99,0.88,0.63}.899 & \cellcolor[rgb]{1.00,0.65,0.45}.670 & \cellcolor[rgb]{1.00,0.78,0.55}.849 & \cellcolor[rgb]{1.00,0.81,0.57}.869 & \cellcolor[rgb]{0.99,0.53,0.38}.600 & \cellcolor[rgb]{1.00,0.78,0.55}.847 & \cellcolor[rgb]{1.00,0.81,0.57}.867 & \cellcolor[rgb]{0.75,0.23,0.46}.336 & \cellcolor[rgb]{0.64,0.19,0.49}.457 & \cellcolor[rgb]{0.79,0.24,0.45}.569 & \cellcolor[rgb]{0.64,0.19,0.49}.261 & \cellcolor[rgb]{0.77,0.23,0.46}.527 & \cellcolor[rgb]{0.87,0.29,0.41}.620 \\
 & Mel/PASE+           & \cellcolor[rgb]{0.99,0.53,0.38}.599 & \cellcolor[rgb]{0.99,0.86,0.62}.885 & \cellcolor[rgb]{0.99,0.86,0.62}.891 & \cellcolor[rgb]{0.98,0.51,0.37}.589 & \cellcolor[rgb]{1.00,0.73,0.51}.821 & \cellcolor[rgb]{1.00,0.68,0.47}.810 & \cellcolor[rgb]{0.98,0.47,0.36}.561 & \cellcolor[rgb]{1.00,0.71,0.49}.812 & \cellcolor[rgb]{1.00,0.74,0.51}.837 & \cellcolor[rgb]{0.97,0.45,0.36}.551 & \cellcolor[rgb]{1.00,0.80,0.56}.852 & \cellcolor[rgb]{1.00,0.81,0.57}.869 & \cellcolor[rgb]{0.66,0.20,0.49}.281 & \cellcolor[rgb]{0.61,0.18,0.50}.440 & \cellcolor[rgb]{0.73,0.22,0.47}.540 & \cellcolor[rgb]{0.66,0.20,0.49}.275 & \cellcolor[rgb]{0.80,0.25,0.44}.549 & \cellcolor[rgb]{0.87,0.29,0.41}.622 \\
 & OpenSmile/PASE+       & \cellcolor[rgb]{0.97,0.45,0.36}.547 & \cellcolor[rgb]{0.99,0.86,0.61}.878 & \cellcolor[rgb]{0.99,0.86,0.62}.892 & \cellcolor[rgb]{0.97,0.44,0.36}.542 & \cellcolor[rgb]{1.00,0.81,0.57}.860 & \cellcolor[rgb]{1.00,0.80,0.56}.861 & \cellcolor[rgb]{0.96,0.40,0.36}.520 & \cellcolor[rgb]{1.00,0.72,0.49}.814 & \cellcolor[rgb]{1.00,0.73,0.51}.830 & \cellcolor[rgb]{0.96,0.43,0.36}.536 & \cellcolor[rgb]{1.00,0.80,0.56}.852 & \cellcolor[rgb]{1.00,0.78,0.55}.856 & \cellcolor[rgb]{0.75,0.23,0.46}.337 & \cellcolor[rgb]{0.75,0.23,0.46}.519 & \cellcolor[rgb]{0.79,0.24,0.45}.576 & \cellcolor[rgb]{0.72,0.21,0.48}.318 & \cellcolor[rgb]{0.82,0.26,0.43}.556 & \cellcolor[rgb]{0.79,0.24,0.45}.570 \\
 & Mel/OpenSmile/Pase+   & \cellcolor[rgb]{0.98,0.51,0.37}.587 & \cellcolor[rgb]{0.99,0.84,0.60}.874 & \cellcolor[rgb]{0.99,0.86,0.61}.888 & \cellcolor[rgb]{0.99,0.57,0.40}.626 & \cellcolor[rgb]{0.99,0.88,0.63}.892 & \cellcolor[rgb]{0.99,0.88,0.63}.900 & \cellcolor[rgb]{0.98,0.51,0.37}.592 & \cellcolor[rgb]{1.00,0.74,0.51}.826 & \cellcolor[rgb]{1.00,0.76,0.52}.842 & \cellcolor[rgb]{0.98,0.49,0.37}.574 & \cellcolor[rgb]{1.00,0.80,0.56}.852 & \cellcolor[rgb]{1.00,0.78,0.55}.859 & \cellcolor[rgb]{0.79,0.24,0.45}.373 & \cellcolor[rgb]{0.79,0.24,0.45}.535 & \cellcolor[rgb]{0.75,0.23,0.46}.555 & \cellcolor[rgb]{0.77,0.24,0.46}.362 & \cellcolor[rgb]{0.87,0.29,0.41}.590 & \cellcolor[rgb]{0.82,0.26,0.43}.591 \\
\midrule
\multirow{7}{*}{DTCNN}
 & Mel-Filterbank      & \cellcolor[rgb]{0.99,0.55,0.39}.612 & \cellcolor[rgb]{0.99,0.86,0.61}.878 & \cellcolor[rgb]{0.99,0.88,0.63}.900 & \cellcolor[rgb]{0.97,0.45,0.36}.550 & \cellcolor[rgb]{1.00,0.72,0.49}.814 & \cellcolor[rgb]{1.00,0.74,0.51}.835 & \cellcolor[rgb]{0.97,0.46,0.36}.553 & \cellcolor[rgb]{1.00,0.73,0.51}.818 & \cellcolor[rgb]{1.00,0.76,0.53}.848 & \cellcolor[rgb]{0.94,0.37,0.37}.504 & \cellcolor[rgb]{1.00,0.76,0.53}.839 & \cellcolor[rgb]{1.00,0.80,0.56}.864 & \cellcolor[rgb]{0.85,0.28,0.42}.410 & \cellcolor[rgb]{0.81,0.25,0.44}.553 & \cellcolor[rgb]{0.77,0.24,0.46}.568 & \cellcolor[rgb]{0.70,0.21,0.48}.313 & \cellcolor[rgb]{0.84,0.27,0.42}.567 & \cellcolor[rgb]{0.77,0.24,0.46}.567 \\
 & OpenSmile             & \cellcolor[rgb]{1.00,0.71,0.49}.703 & \cellcolor[rgb]{0.99,0.91,0.66}.908 & \cellcolor[rgb]{0.99,0.93,0.68}.924 & \cellcolor[rgb]{1.00,0.64,0.44}.666 & \cellcolor[rgb]{0.99,0.93,0.68}.916 & \cellcolor[rgb]{0.99,0.94,0.70}.928 & \cellcolor[rgb]{0.99,0.59,0.41}.640 & \cellcolor[rgb]{1.00,0.81,0.57}.856 & \cellcolor[rgb]{1.00,0.82,0.58}.873 & \cellcolor[rgb]{0.99,0.57,0.40}.618 & \cellcolor[rgb]{0.99,0.84,0.60}.876 & \cellcolor[rgb]{0.99,0.86,0.61}.888 & \cellcolor[rgb]{0.90,0.31,0.39}.452 & \cellcolor[rgb]{0.86,0.28,0.42}.583 & \cellcolor[rgb]{0.87,0.29,0.41}.614 & \cellcolor[rgb]{0.79,0.24,0.45}.375 & \cellcolor[rgb]{0.93,0.35,0.37}.634 & \cellcolor[rgb]{0.94,0.36,0.37}.664 \\
 & PASE+               & \cellcolor[rgb]{1.00,0.80,0.56}.759 & \cellcolor[rgb]{0.99,0.88,0.63}.893 & \cellcolor[rgb]{0.99,0.88,0.64}.901 & \cellcolor[rgb]{1.00,0.63,0.43}.658 & \cellcolor[rgb]{1.00,0.68,0.47}.801 & \cellcolor[rgb]{1.00,0.68,0.46}.806 & \cellcolor[rgb]{1.00,0.72,0.49}.711 & \cellcolor[rgb]{1.00,0.78,0.55}.846 & \cellcolor[rgb]{1.00,0.78,0.54}.851 & \cellcolor[rgb]{0.99,0.59,0.41}.628 & \cellcolor[rgb]{0.99,0.59,0.41}.757 & \cellcolor[rgb]{0.99,0.57,0.40}.762 & \cellcolor[rgb]{0.94,0.36,0.37}.496 & \cellcolor[rgb]{0.87,0.29,0.41}.594 & \cellcolor[rgb]{0.85,0.28,0.42}.604 & \cellcolor[rgb]{0.80,0.25,0.44}.379 & \cellcolor[rgb]{0.77,0.24,0.46}.531 & \cellcolor[rgb]{0.68,0.20,0.49}.514 \\
 & Mel/OpenSmile         & \cellcolor[rgb]{0.99,0.90,0.65}.824 & \cellcolor[rgb]{0.99,0.96,0.72}.934 & \cellcolor[rgb]{0.99,0.95,0.70}.931 & \cellcolor[rgb]{1.00,0.72,0.49}.710 & \cellcolor[rgb]{0.99,0.91,0.66}.907 & \cellcolor[rgb]{0.99,0.88,0.63}.897 & \cellcolor[rgb]{1.00,0.76,0.53}.739 & \cellcolor[rgb]{0.99,0.88,0.64}.897 & \cellcolor[rgb]{0.99,0.88,0.63}.897 & \cellcolor[rgb]{1.00,0.71,0.49}.709 & \cellcolor[rgb]{0.99,0.91,0.66}.908 & \cellcolor[rgb]{0.99,0.90,0.65}.908 & \cellcolor[rgb]{0.84,0.27,0.42}.407 & \cellcolor[rgb]{0.79,0.24,0.45}.536 & \cellcolor[rgb]{0.81,0.25,0.44}.585 & \cellcolor[rgb]{0.77,0.23,0.46}.353 & \cellcolor[rgb]{0.91,0.33,0.38}.616 & \cellcolor[rgb]{0.95,0.38,0.36}.674 \\
 & Mel/PASE+           & \cellcolor[rgb]{1.00,0.78,0.55}.752 & \cellcolor[rgb]{0.99,0.92,0.68}.913 & \cellcolor[rgb]{0.99,0.91,0.66}.914 & \cellcolor[rgb]{0.99,0.59,0.41}.638 & \cellcolor[rgb]{1.00,0.80,0.56}.854 & \cellcolor[rgb]{1.00,0.76,0.53}.847 & \cellcolor[rgb]{1.00,0.72,0.49}.715 & \cellcolor[rgb]{0.99,0.84,0.60}.872 & \cellcolor[rgb]{1.00,0.82,0.58}.873 & \cellcolor[rgb]{0.95,0.38,0.36}.509 & \cellcolor[rgb]{1.00,0.71,0.49}.811 & \cellcolor[rgb]{1.00,0.64,0.44}.791 & \cellcolor[rgb]{0.90,0.31,0.39}.455 & \cellcolor[rgb]{0.84,0.27,0.42}.572 & \cellcolor[rgb]{0.83,0.26,0.43}.595 & \cellcolor[rgb]{0.79,0.24,0.45}.373 & \cellcolor[rgb]{0.68,0.20,0.49}.478 & \cellcolor[rgb]{0.59,0.17,0.50}.472 \\
 & OpenSmile/PASE+       & \cellcolor[rgb]{1.00,0.63,0.43}.661 & \cellcolor[rgb]{1.00,0.83,0.59}.870 & \cellcolor[rgb]{1.00,0.83,0.59}.877 & \cellcolor[rgb]{0.99,0.54,0.39}.602 & \cellcolor[rgb]{0.99,0.86,0.62}.888 & \cellcolor[rgb]{0.99,0.84,0.60}.885 & \cellcolor[rgb]{1.00,0.78,0.55}.755 & \cellcolor[rgb]{0.99,0.86,0.61}.878 & \cellcolor[rgb]{1.00,0.83,0.59}.877 & \cellcolor[rgb]{0.99,0.55,0.39}.610 & \cellcolor[rgb]{1.00,0.68,0.47}.798 & \cellcolor[rgb]{1.00,0.64,0.44}.792 & \cellcolor[rgb]{0.90,0.32,0.39}.459 & \cellcolor[rgb]{0.84,0.27,0.42}.571 & \cellcolor[rgb]{0.79,0.24,0.45}.570 & \cellcolor[rgb]{0.75,0.23,0.46}.335 & \cellcolor[rgb]{0.68,0.20,0.49}.479 & \cellcolor[rgb]{0.55,0.16,0.51}.450 \\
 & Mel/OpenSmile/Pase+   & \cellcolor[rgb]{0.99,0.86,0.61}.792 & \cellcolor[rgb]{0.99,0.93,0.68}.918 & \cellcolor[rgb]{0.99,0.92,0.68}.919 & \cellcolor[rgb]{1.00,0.65,0.45}.673 & \cellcolor[rgb]{1.00,0.73,0.51}.822 & \cellcolor[rgb]{1.00,0.68,0.47}.812 & \cellcolor[rgb]{1.00,0.73,0.51}.719 & \cellcolor[rgb]{0.99,0.86,0.62}.887 & \cellcolor[rgb]{0.99,0.86,0.61}.887 & \cellcolor[rgb]{0.97,0.46,0.36}.553 & \cellcolor[rgb]{1.00,0.68,0.47}.796 & \cellcolor[rgb]{1.00,0.64,0.44}.791 & \cellcolor[rgb]{0.94,0.36,0.37}.498 & \cellcolor[rgb]{0.83,0.26,0.43}.565 & \cellcolor[rgb]{0.84,0.27,0.42}.601 & \cellcolor[rgb]{0.85,0.28,0.42}.413 & \cellcolor[rgb]{0.70,0.21,0.48}.490 & \cellcolor[rgb]{0.66,0.20,0.49}.503 \\
\midrule
\multirow{7}{*}{GRU}
 & \underline{Mel-Filterbank}     & \cellcolor[rgb]{1.00,0.78,0.55}.755 & \cellcolor[rgb]{0.99,0.91,0.66}.906 & \cellcolor[rgb]{0.99,0.91,0.66}.913 & \cellcolor[rgb]{0.99,0.59,0.41}.634 & \cellcolor[rgb]{0.99,0.88,0.64}.897 & \cellcolor[rgb]{0.99,0.88,0.64}.904 & \cellcolor[rgb]{0.99,0.61,0.42}.644 & \cellcolor[rgb]{1.00,0.76,0.53}.837 & \cellcolor[rgb]{1.00,0.78,0.55}.856 & \cellcolor[rgb]{0.96,0.43,0.36}.536 & \cellcolor[rgb]{1.00,0.76,0.52}.833 & \cellcolor[rgb]{1.00,0.78,0.54}.851 & \cellcolor[rgb]{0.85,0.28,0.42}.416 & \cellcolor[rgb]{0.77,0.23,0.46}.528 & \cellcolor[rgb]{0.77,0.24,0.46}.567 & \cellcolor[rgb]{0.75,0.23,0.46}.344 & \cellcolor[rgb]{0.87,0.29,0.41}.589 & \cellcolor[rgb]{0.90,0.32,0.39}.643 \\
 & OpenSmile             & \cellcolor[rgb]{1.00,0.76,0.53}.738 & \cellcolor[rgb]{0.99,0.93,0.68}.921 & \cellcolor[rgb]{0.99,0.96,0.72}.937 & \cellcolor[rgb]{1.00,0.64,0.44}.663 & \cellcolor[rgb]{0.99,0.91,0.66}.908 & \cellcolor[rgb]{0.99,0.92,0.68}.919 & \cellcolor[rgb]{1.00,0.74,0.51}.723 & \cellcolor[rgb]{0.99,0.86,0.62}.885 & \cellcolor[rgb]{0.99,0.88,0.64}.903 & \cellcolor[rgb]{0.99,0.61,0.42}.646 & \cellcolor[rgb]{0.99,0.86,0.61}.881 & \cellcolor[rgb]{0.99,0.86,0.62}.894 & \cellcolor[rgb]{0.89,0.31,0.40}.448 & \cellcolor[rgb]{0.81,0.25,0.44}.554 & \cellcolor[rgb]{0.85,0.28,0.42}.604 & \cellcolor[rgb]{0.81,0.25,0.44}.385 & \cellcolor[rgb]{0.89,0.31,0.40}.603 & \cellcolor[rgb]{0.90,0.31,0.39}.634 \\
 & PASE+               & \cellcolor[rgb]{1.00,0.74,0.51}.726 & \cellcolor[rgb]{0.99,0.93,0.68}.919 & \cellcolor[rgb]{0.99,0.94,0.70}.927 & \cellcolor[rgb]{1.00,0.73,0.51}.717 & \cellcolor[rgb]{0.99,0.93,0.68}.916 & \cellcolor[rgb]{0.99,0.94,0.70}.928 & \cellcolor[rgb]{1.00,0.76,0.52}.731 & \cellcolor[rgb]{0.99,0.88,0.64}.895 & \cellcolor[rgb]{0.99,0.90,0.65}.908 & \cellcolor[rgb]{1.00,0.64,0.44}.664 & \cellcolor[rgb]{0.99,0.90,0.65}.900 & \cellcolor[rgb]{0.99,0.91,0.66}.912 & \cellcolor[rgb]{0.96,0.40,0.36}.521 & \cellcolor[rgb]{0.93,0.35,0.37}.635 & \cellcolor[rgb]{0.94,0.36,0.37}.664 & \cellcolor[rgb]{0.90,0.31,0.39}.452 & \cellcolor[rgb]{0.96,0.40,0.36}.661 & \cellcolor[rgb]{0.97,0.44,0.36}.700 \\
 & Mel/OpenSmile         & \cellcolor[rgb]{1.00,0.82,0.58}.774 & \cellcolor[rgb]{0.99,0.94,0.70}.924 & \cellcolor[rgb]{0.99,0.95,0.70}.932 & \cellcolor[rgb]{1.00,0.66,0.45}.679 & \cellcolor[rgb]{0.99,0.92,0.68}.912 & \cellcolor[rgb]{0.99,0.92,0.68}.920 & \cellcolor[rgb]{1.00,0.78,0.54}.745 & \cellcolor[rgb]{0.99,0.88,0.63}.891 & \cellcolor[rgb]{0.99,0.88,0.63}.900 & \cellcolor[rgb]{1.00,0.68,0.46}.684 & \cellcolor[rgb]{0.99,0.88,0.64}.895 & \cellcolor[rgb]{0.99,0.88,0.64}.902 & \cellcolor[rgb]{0.96,0.40,0.36}.523 & \cellcolor[rgb]{0.88,0.30,0.40}.595 & \cellcolor[rgb]{0.88,0.30,0.40}.625 & \cellcolor[rgb]{0.90,0.31,0.39}.454 & \cellcolor[rgb]{0.93,0.35,0.37}.636 & \cellcolor[rgb]{0.93,0.35,0.37}.656 \\
 & Mel/PASE+           & \cellcolor[rgb]{1.00,0.74,0.51}.725 & \cellcolor[rgb]{0.99,0.94,0.70}.924 & \cellcolor[rgb]{0.99,0.94,0.70}.930 & \cellcolor[rgb]{0.99,0.59,0.41}.640 & \cellcolor[rgb]{0.99,0.91,0.66}.909 & \cellcolor[rgb]{0.99,0.92,0.68}.917 & \cellcolor[rgb]{1.00,0.74,0.51}.726 & \cellcolor[rgb]{0.99,0.88,0.64}.899 & \cellcolor[rgb]{0.99,0.90,0.65}.909 & \cellcolor[rgb]{1.00,0.65,0.45}.674 & \cellcolor[rgb]{0.99,0.90,0.65}.903 & \cellcolor[rgb]{0.99,0.91,0.66}.912 & \cellcolor[rgb]{0.97,0.45,0.36}.546 & \cellcolor[rgb]{0.93,0.35,0.37}.635 & \cellcolor[rgb]{0.91,0.33,0.38}.644 & \cellcolor[rgb]{0.93,0.35,0.37}.482 & \cellcolor[rgb]{0.96,0.43,0.36}.672 & \cellcolor[rgb]{0.96,0.40,0.36}.687 \\
 & OpenSmile/PASE+       & \cellcolor[rgb]{1.00,0.83,0.59}.778 & \cellcolor[rgb]{0.99,0.94,0.70}.926 & \cellcolor[rgb]{0.99,0.95,0.70}.935 & \cellcolor[rgb]{1.00,0.74,0.51}.729 & \cellcolor[rgb]{0.99,0.93,0.68}.916 & \cellcolor[rgb]{0.99,0.94,0.70}.929 & \cellcolor[rgb]{1.00,0.78,0.55}.752 & \cellcolor[rgb]{0.99,0.88,0.64}.899 & \cellcolor[rgb]{0.99,0.90,0.65}.909 & \cellcolor[rgb]{0.99,0.59,0.41}.632 & \cellcolor[rgb]{0.99,0.88,0.64}.897 & \cellcolor[rgb]{0.99,0.90,0.65}.909 & \cellcolor[rgb]{0.98,0.48,0.37}.573 & \cellcolor[rgb]{0.95,0.38,0.36}.650 & \cellcolor[rgb]{0.93,0.35,0.37}.655 & \cellcolor[rgb]{0.93,0.35,0.37}.485 & \cellcolor[rgb]{0.94,0.36,0.37}.642 & \cellcolor[rgb]{0.90,0.32,0.39}.640 \\
 & Mel/OpenSmile/Pase+   & \cellcolor[rgb]{0.99,0.88,0.64}.813 & \cellcolor[rgb]{0.99,0.95,0.70}.929 & \cellcolor[rgb]{0.99,0.95,0.70}.935 & \cellcolor[rgb]{1.00,0.76,0.53}.742 & \cellcolor[rgb]{0.99,0.94,0.70}.922 & \cellcolor[rgb]{0.99,0.95,0.70}.931 & \cellcolor[rgb]{1.00,0.82,0.58}.772 & \cellcolor[rgb]{0.99,0.91,0.66}.907 & \cellcolor[rgb]{0.99,0.91,0.66}.914 & \cellcolor[rgb]{1.00,0.76,0.52}.733 & \cellcolor[rgb]{0.99,0.93,0.68}\textbf{.916} & \cellcolor[rgb]{0.99,0.93,0.68}.921 & \cellcolor[rgb]{0.96,0.43,0.36}.533 & \cellcolor[rgb]{0.92,0.34,0.38}.626 & \cellcolor[rgb]{0.92,0.34,0.38}.651 & \cellcolor[rgb]{0.93,0.35,0.37}.490 & \cellcolor[rgb]{0.94,0.36,0.37}.638 & \cellcolor[rgb]{0.92,0.34,0.38}.652 \\
\midrule
\multirow{7}{*}{BiGRU}
 & Mel-Filterbank      & \cellcolor[rgb]{1.00,0.70,0.48}.697 & \cellcolor[rgb]{0.99,0.91,0.66}.909 & \cellcolor[rgb]{0.99,0.92,0.68}.917 & \cellcolor[rgb]{1.00,0.63,0.43}.660 & \cellcolor[rgb]{1.00,0.78,0.54}.843 & \cellcolor[rgb]{1.00,0.78,0.55}.855 & \cellcolor[rgb]{1.00,0.78,0.54}.748 & \cellcolor[rgb]{0.99,0.86,0.62}.883 & \cellcolor[rgb]{0.99,0.88,0.63}.896 & \cellcolor[rgb]{0.99,0.59,0.41}.630 & \cellcolor[rgb]{0.99,0.86,0.61}.878 & \cellcolor[rgb]{0.99,0.86,0.62}.891 & \cellcolor[rgb]{0.94,0.36,0.37}.492 & \cellcolor[rgb]{0.88,0.30,0.40}.599 & \cellcolor[rgb]{0.88,0.30,0.40}.628 & \cellcolor[rgb]{0.75,0.23,0.46}.338 & \cellcolor[rgb]{0.83,0.26,0.43}.565 & \cellcolor[rgb]{0.81,0.25,0.44}.584 \\
 & OpenSmile             & \cellcolor[rgb]{1.00,0.76,0.53}.741 & \cellcolor[rgb]{0.99,0.93,0.68}.920 & \cellcolor[rgb]{0.99,0.96,0.72}.937 & \cellcolor[rgb]{1.00,0.78,0.54}\textbf{.748} & \cellcolor[rgb]{0.99,0.94,0.70}.922 & \cellcolor[rgb]{0.99,0.96,0.72}.936 & \cellcolor[rgb]{1.00,0.78,0.54}.748 & \cellcolor[rgb]{0.99,0.88,0.64}.895 & \cellcolor[rgb]{0.99,0.90,0.65}.909 & \cellcolor[rgb]{1.00,0.76,0.53}.738 & \cellcolor[rgb]{0.99,0.91,0.66}.905 & \cellcolor[rgb]{0.99,0.91,0.66}.915 & \cellcolor[rgb]{0.90,0.32,0.39}.461 & \cellcolor[rgb]{0.81,0.25,0.44}.553 & \cellcolor[rgb]{0.83,0.26,0.43}.596 & \cellcolor[rgb]{0.87,0.29,0.41}.436 & \cellcolor[rgb]{0.87,0.29,0.41}.591 & \cellcolor[rgb]{0.84,0.27,0.42}.601 \\
 & PASE+               & \cellcolor[rgb]{0.99,0.91,0.66}.831 & \cellcolor[rgb]{0.99,0.97,0.73}.939 & \cellcolor[rgb]{0.99,0.98,0.74}.946 & \cellcolor[rgb]{1.00,0.70,0.48}.696 & \cellcolor[rgb]{0.99,0.94,0.70}.925 & \cellcolor[rgb]{0.99,0.96,0.72}\textbf{.938} & \cellcolor[rgb]{0.99,0.96,0.72}.860 & \cellcolor[rgb]{0.99,0.95,0.70}.930 & \cellcolor[rgb]{0.99,0.95,0.70}.931 & \cellcolor[rgb]{1.00,0.70,0.48}.701 & \cellcolor[rgb]{0.99,0.88,0.63}.891 & \cellcolor[rgb]{0.99,0.84,0.60}.882 & \cellcolor[rgb]{1.00,0.63,0.43}\textbf{.655} & \cellcolor[rgb]{0.97,0.46,0.36}\textbf{.688} & \cellcolor[rgb]{0.96,0.43,0.36}.698 & \cellcolor[rgb]{0.97,0.44,0.36}.541 & \cellcolor[rgb]{0.98,0.48,0.37}.701 & \cellcolor[rgb]{0.97,0.45,0.36}.707 \\
 & Mel/OpenSmile         & \cellcolor[rgb]{0.99,0.90,0.65}.820 & \cellcolor[rgb]{0.99,0.96,0.72}.937 & \cellcolor[rgb]{0.99,0.97,0.73}.941 & \cellcolor[rgb]{1.00,0.76,0.53}.742 & \cellcolor[rgb]{0.99,0.95,0.70}\textbf{.928} & \cellcolor[rgb]{0.99,0.95,0.70}.935 & \cellcolor[rgb]{0.99,0.88,0.63}.809 & \cellcolor[rgb]{0.99,0.90,0.65}.902 & \cellcolor[rgb]{0.99,0.88,0.64}.902 & \cellcolor[rgb]{1.00,0.76,0.53}\textbf{.743} & \cellcolor[rgb]{0.99,0.91,0.66}.910 & \cellcolor[rgb]{0.99,0.91,0.66}.911 & \cellcolor[rgb]{0.97,0.46,0.36}.556 & \cellcolor[rgb]{0.90,0.32,0.39}.613 & \cellcolor[rgb]{0.91,0.33,0.38}.647 & \cellcolor[rgb]{0.90,0.32,0.39}.460 & \cellcolor[rgb]{0.95,0.38,0.36}.652 & \cellcolor[rgb]{0.94,0.36,0.37}.667 \\
 & Mel/PASE+           & \cellcolor[rgb]{0.99,0.94,0.70}.846 & \cellcolor[rgb]{0.99,0.99,0.75}\textbf{.951} & \cellcolor[rgb]{0.99,0.99,0.75}.952 & \cellcolor[rgb]{1.00,0.64,0.44}.667 & \cellcolor[rgb]{0.99,0.91,0.66}.907 & \cellcolor[rgb]{0.99,0.88,0.64}.902 & \cellcolor[rgb]{0.99,0.95,0.70}.856 & \cellcolor[rgb]{0.99,0.95,0.70}.929 & \cellcolor[rgb]{0.99,0.95,0.70}.933 & \cellcolor[rgb]{1.00,0.71,0.49}.709 & \cellcolor[rgb]{0.99,0.91,0.66}.908 & \cellcolor[rgb]{0.99,0.92,0.68}.919 & \cellcolor[rgb]{0.99,0.61,0.42}.645 & \cellcolor[rgb]{0.97,0.44,0.36}.676 & \cellcolor[rgb]{0.96,0.40,0.36}.685 & \cellcolor[rgb]{0.95,0.38,0.36}.509 & \cellcolor[rgb]{0.95,0.40,0.36}.656 & \cellcolor[rgb]{0.93,0.35,0.37}.660 \\
 & OpenSmile/PASE+       & \cellcolor[rgb]{0.99,0.90,0.65}.823 & \cellcolor[rgb]{0.99,0.97,0.73}.941 & \cellcolor[rgb]{0.99,0.97,0.73}.945 & \cellcolor[rgb]{1.00,0.66,0.45}.680 & \cellcolor[rgb]{0.99,0.94,0.70}.923 & \cellcolor[rgb]{0.99,0.94,0.70}.928 & \cellcolor[rgb]{0.99,0.94,0.70}.847 & \cellcolor[rgb]{0.99,0.94,0.70}.922 & \cellcolor[rgb]{0.99,0.94,0.70}.928 & \cellcolor[rgb]{1.00,0.71,0.49}.708 & \cellcolor[rgb]{0.99,0.90,0.65}.901 & \cellcolor[rgb]{0.99,0.91,0.66}.912 & \cellcolor[rgb]{0.99,0.55,0.39}.614 & \cellcolor[rgb]{0.97,0.44,0.36}.677 & \cellcolor[rgb]{0.97,0.44,0.36}\textbf{.701} & \cellcolor[rgb]{0.94,0.37,0.37}.500 & \cellcolor[rgb]{0.97,0.46,0.36}.689 & \cellcolor[rgb]{0.98,0.48,0.37}\textbf{.720} \\
 & Mel/OpenSmile/Pase+   & \cellcolor[rgb]{0.99,0.99,0.75}\textbf{.883} & \cellcolor[rgb]{0.99,0.99,0.75}\textbf{.951} & \cellcolor[rgb]{0.99,0.99,0.75}\textbf{.953} & \cellcolor[rgb]{1.00,0.66,0.45}.681 & \cellcolor[rgb]{0.99,0.92,0.68}.914 & \cellcolor[rgb]{0.99,0.91,0.66}.912 & \cellcolor[rgb]{0.99,0.98,0.74}\textbf{.876} & \cellcolor[rgb]{0.99,0.96,0.72}\textbf{.935} & \cellcolor[rgb]{0.99,0.96,0.72}\textbf{.936} & \cellcolor[rgb]{1.00,0.73,0.51}.717 & \cellcolor[rgb]{0.99,0.93,0.68}\textbf{.916} & \cellcolor[rgb]{0.99,0.93,0.68}\textbf{.923} & \cellcolor[rgb]{0.99,0.59,0.41}.640 & \cellcolor[rgb]{0.96,0.43,0.36}.671 & \cellcolor[rgb]{0.95,0.38,0.36}.678 & \cellcolor[rgb]{0.98,0.47,0.36}\textbf{.565} & \cellcolor[rgb]{0.98,0.49,0.37}\textbf{.706} & \cellcolor[rgb]{0.98,0.47,0.36}.715 \\
\bottomrule
\end{tabular}%
}
\end{table*}

\subsection{Results}

\subsubsection{Model Comparison}
We evaluate each model across single and combined feature sets for each label granularity. The results are shown in Table~\ref{tab:main-results} which reports mean average precision (mAP), class frequency weighted $F_1$-score (F) and accuracy (Acc). This table also includes a baseline performance that uses the training set's empirical class label distribution as the posterior probability of each frame.
We note several observations. 

First, as expected, the difficulty of the task increases with the number of labels in the label set. However, the drop off in performance from 4-way to 5-way, which requires differentiating instructor from student voice, was less pronounced than anticipated.
Second, the BiGRU models outperform every other model for every task (although the GRU is often close behind), and in nearly every case more than doubles the percent increase in performance over baseline from 4-way to 9-way when using all features.
Third, feature combinations that include PASE+ embeddings almost always perform better. However, the combination of mel-filterbank and OpenSmile features performs 4.3\% and 7.7\% better by F on the 4-way DNN classifier than mel-filterbank or prosody alone, respectively, and 7.3\% better than PASE+ embeddings alone.
Our last observation on Table~\ref{tab:main-results} is that there is some degradation in performance from Test 1 (unseen class sessions from previously seen instructors) to Test 2 (unseen class sessions from unseen instructors), particularly in mAP, although by F and Acc the model generalizes comparably. The generally good performance on Test~2 provides evidence that the model can generalize to new instructors, buildings, classrooms, courses and disciplines. We leave to future work additional tests of generalization; e.g., to different universities, microphones or microphone placement strategies.

\subsubsection{Confusion Analysis}
\begin{figure}
  \centering
  \includegraphics[width=0.30\columnwidth]{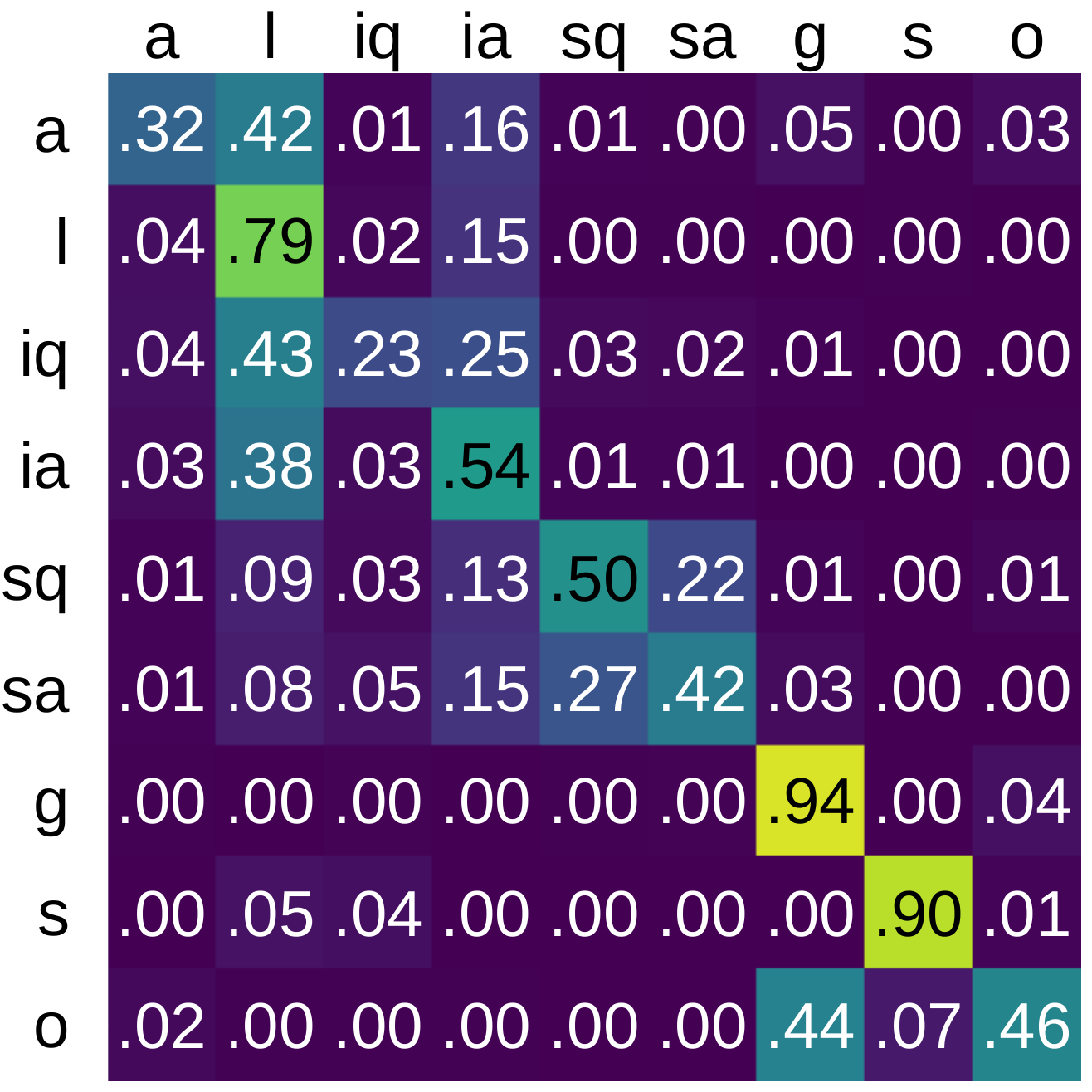}
  \includegraphics[width=0.30\columnwidth]{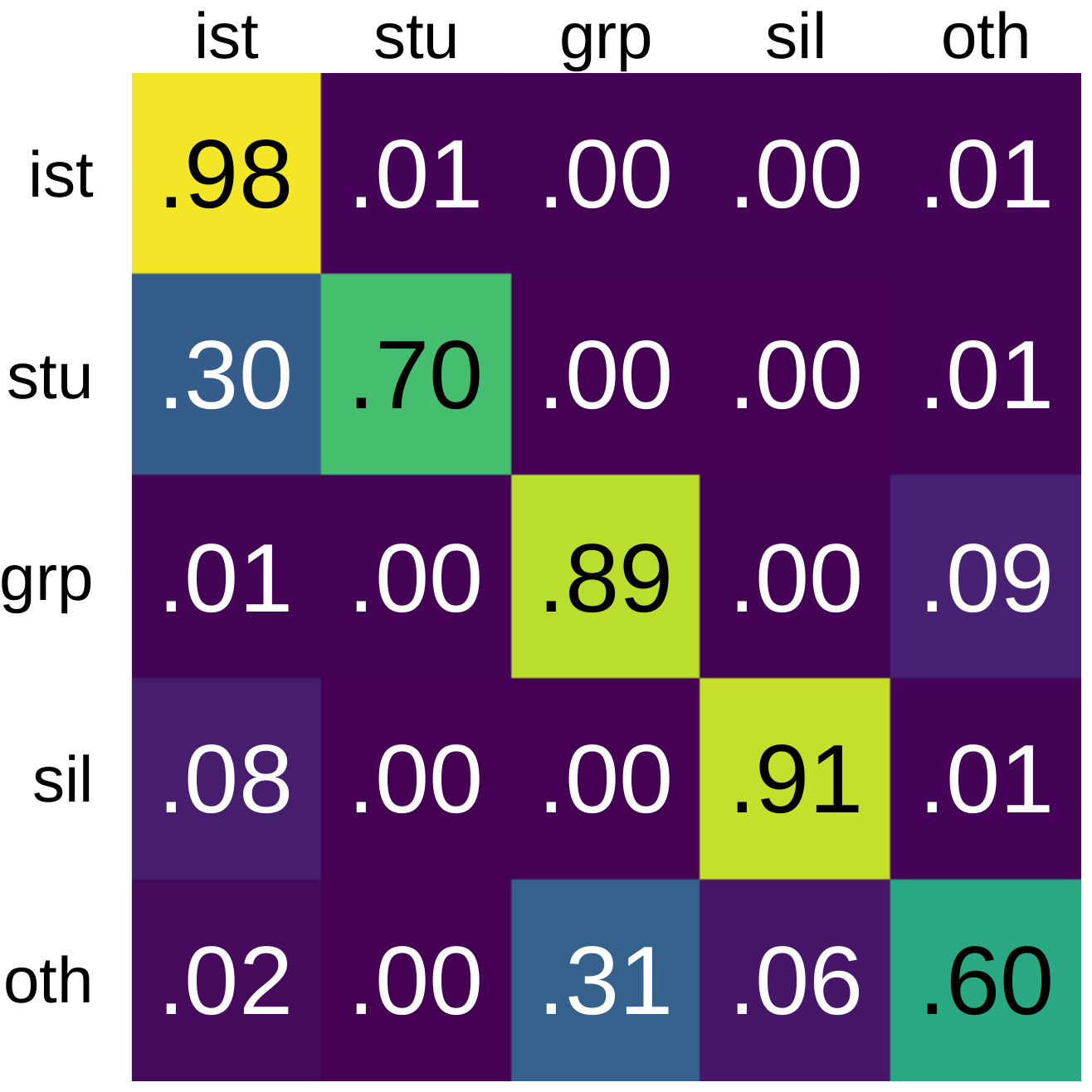}
  \includegraphics[width=0.36\columnwidth]{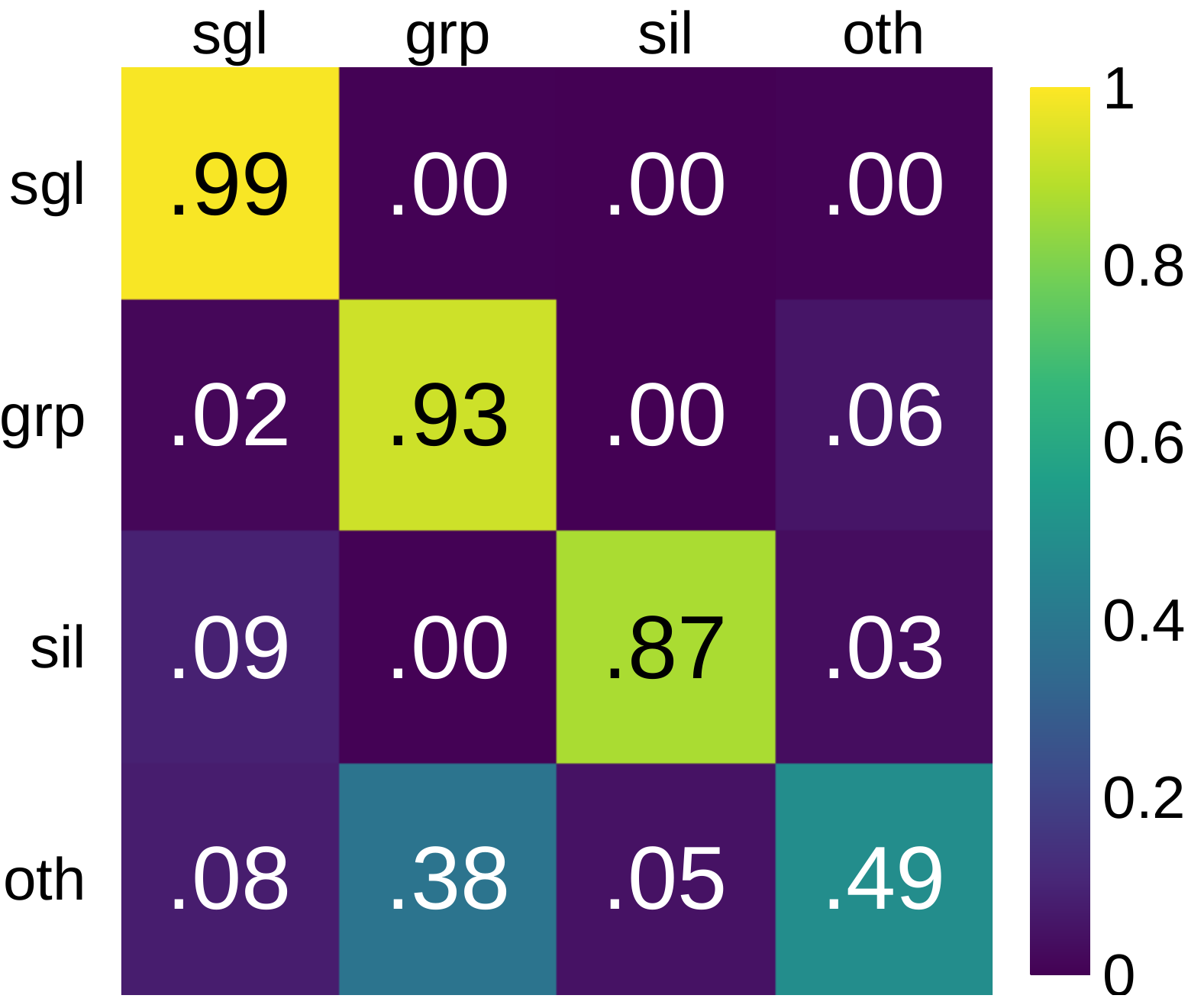}
  \caption{Confusion matrices for the best overall model (BiGRU, all features) performing 9-way, 5-way, and 4-way classification on Test 1, shown from left to right.}
  \label{fig:confusion-matrix}
\end{figure}

To better understand what types of mistakes the model makes, we provide three confusion matrices in Fig.~\ref{fig:confusion-matrix}. On the 9-way matrix we see confusion between lecture and other instructor speech (announcements, instructor questions and answers). This is expected given that these distinctions are largely semantic, but our models use no lexical features. The 5-way confusion matrix shows that students are most often misclassified as instructors, an understandable confusion between the two single voice categories, and that the two most chaotic activities are sometimes confused, with the unstructured ``other'' often misclassified as ``group work.'' Lastly, in the 4-way confusion matrix, we see very strong performance, although some ``group work'' and ``other'' confusion remains.

\subsubsection{Precision-Recall Analysis}
\begin{figure}
  \centering
  \includegraphics[width=0.7\columnwidth]{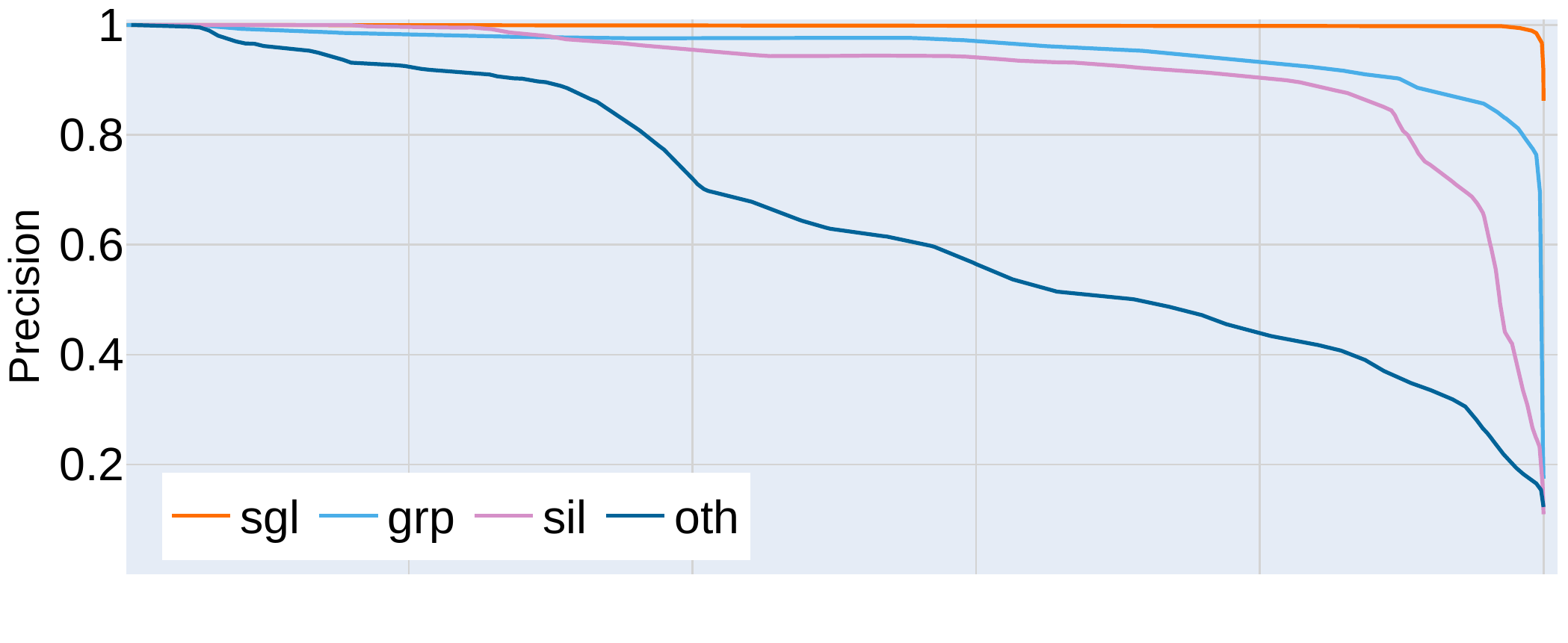}
  \includegraphics[width=0.7\columnwidth]{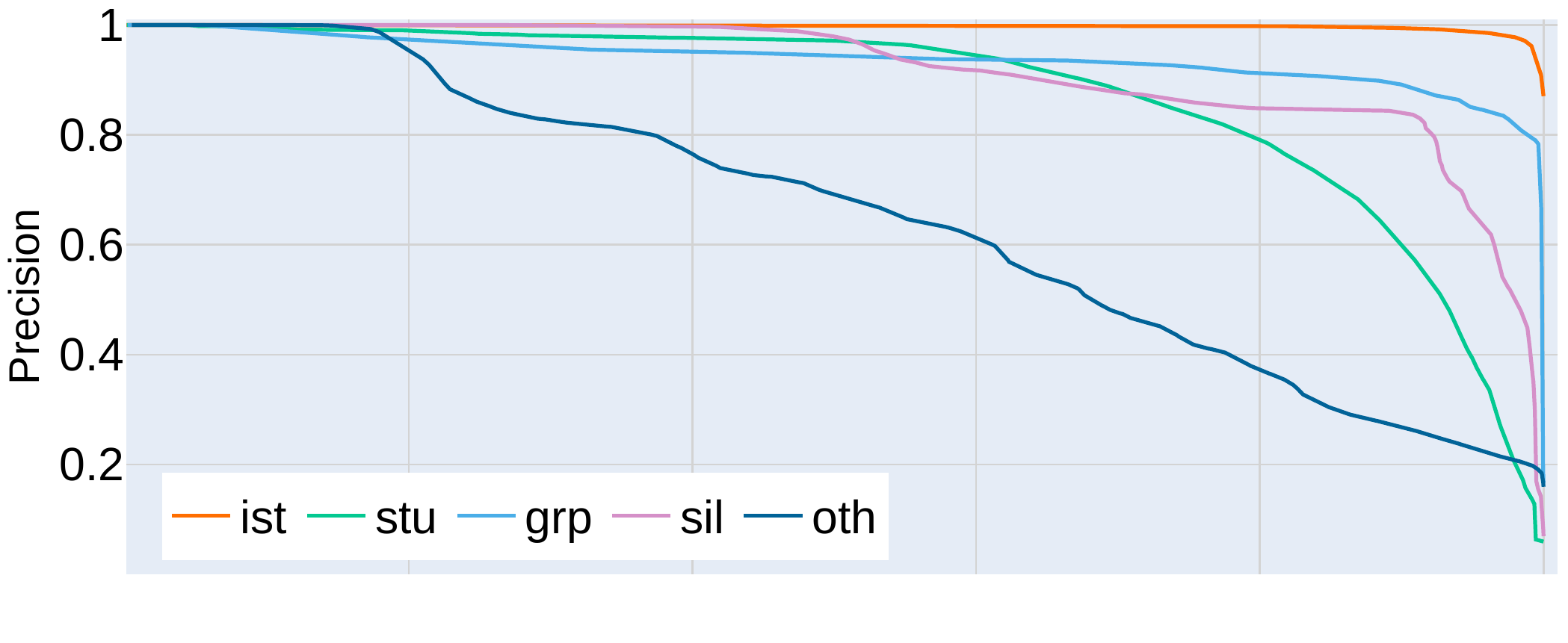}
  \includegraphics[width=0.7\columnwidth]{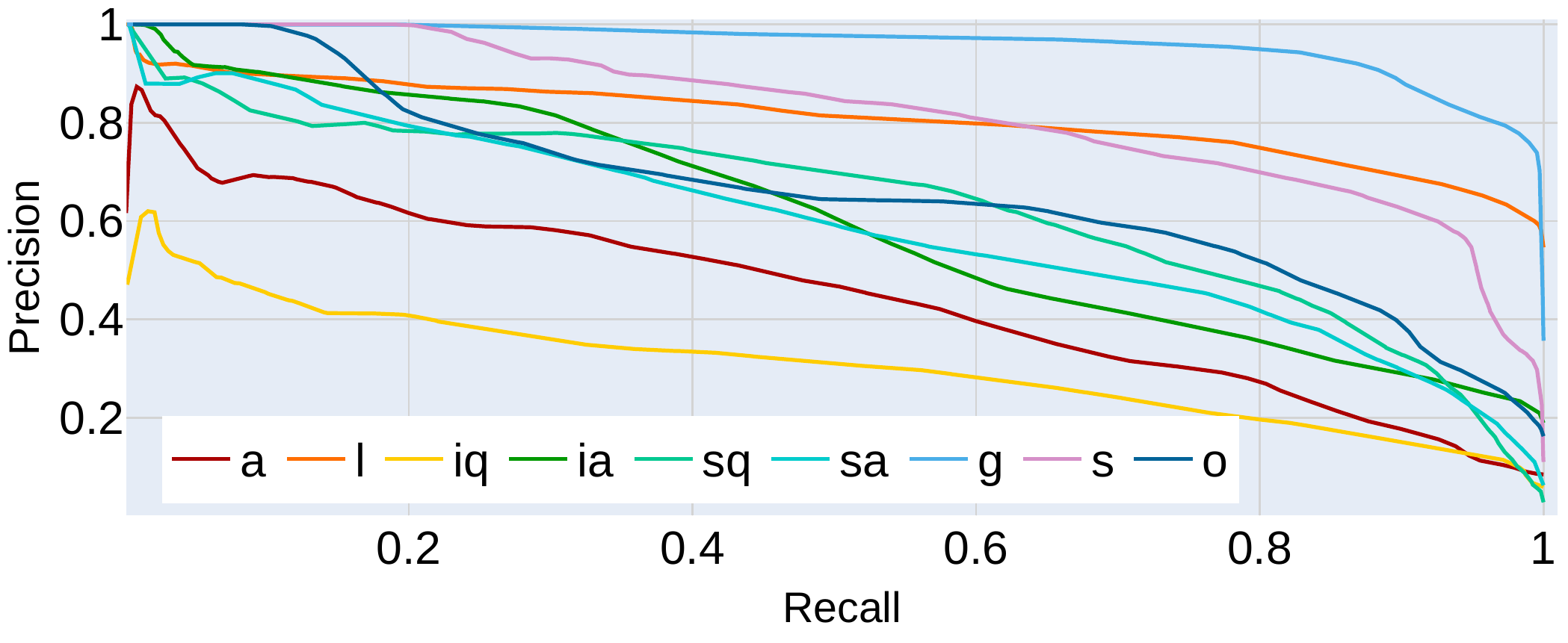}
  \caption{Precision-recall curves for the best overall model (BiGRU, all features) on Test 1, using the 4-way (top), 5-way (middle) and 9-way (bottom) label schemes.}
  \label{fig:pr-curve}
\end{figure}

Precision-recall curves for the best overall model in each labeling scheme are plotted in Fig.~\ref{fig:pr-curve}. All classes besides ``other'' are able to achieve at least 0.8 recall and 0.8 precision simultaneously, with ``single-voice'' (sgl) and ``instructor'' able to achieve near perfect recall and precision in the 4- and 5-way case, respectively.

\subsubsection{Aggregate Time on Activity Analysis}
\begin{figure}
  \centering
  \includegraphics[width=0.9\columnwidth]{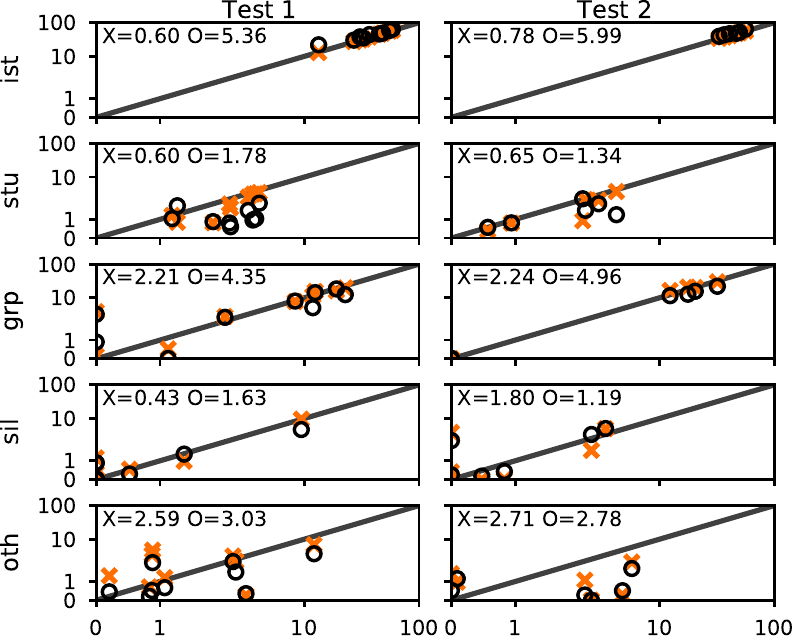}
  \caption{Predicted vs actual (vertical vs horizontal axes) time spent (minutes), by activity, in each Test 1 (left column) or 2 (right column) class session. Orange $X$s denote BiGRU with all features; black $O$s denote the best model reported in \cite{cosbey2019deep}. Root mean squared error (in minutes) between predicted and actual is listed in the upper left of each subfigure.}
  \label{fig:contributions}
\end{figure}

Fig.~\ref{fig:contributions} shows scatter plots comparing the predicted and actual aggregate time spent in each activity for Test 1 and Test 2. Each point represents a class session, while points above (below) the diagonal indicate that the predicted time spent in the event over (under) estimates the true time. Our best model is compared against our strongest baseline, the best model of \cite{cosbey2019deep}. We see dramatically reduced root mean squared error (RMSE, in minutes): 54.9\% lower when averaged uniformly across all categories and both test sets.

\section{Conclusions and Future Work} \label{sec:conclusions}
We present a set of methods for fine-grained classroom activity detection from audio, and extensively evaluate the performance of multiple neural network classifiers, feature sets, and label sets. We find the best performing model to be a BiGRU using mel-filterbank, OpenSmile, and PASE+ features, although other models are competitive. In the majority of cases, self-supervised PASE+ features, alone or combined with other features, yield the best results. The best models report 0.928, 0.916 and 0.706 weighted-$F_1$ for the \mbox{4-,} 5- and 9-way label sets, respectively, when generalizing to previously unseen instructors. Our best model also produces highly accurate estimates of aggregate class time spent on activities: RMSE of 1.3 and 1.6 minutes averaged over activity class labels for the 5-way task when generalizing to unseen class sessions and unseen instructors, respectively -- a 60.2\% and 49.7\% drop in RMSE compared to a strong baseline.

One direction for future work would be to incorporate lexical features from a high quality far-field noise-robust ASR system, which may help differentiate sub-classes of single-speaker audio (e.g., finding characteristic phrases \cite{Zhang10,zhang13} of classroom activities \cite{10.1145/3027385.3027417}). One might consider alternative architectures (e.g., Transformer \cite{vaswani17} variants). Multi-task learning between labels of different granularity may improve performance, further reducing the already small degradation in performance when mapping from fine to coarse-grained predictions. Finally, multi-modal approaches, incorporating both audio and video, may also help to disambiguate by leveraging cross-modal correlations (e.g., hands raising before student question or answer).

\bibliography{refs.bib}{}

\end{document}